\documentclass[useAMS,referee]{mn2e}
\usepackage{subfig,graphicx}
\usepackage{multirow}
\usepackage{amsmath}
\usepackage{lscape}
\usepackage{float}
\usepackage{stfloats}
\usepackage{enumerate}
\usepackage{fixltx2e}
\usepackage{afterpage}

\usepackage{natbib}
\bibpunct{(}{)}{;}{a}{}{,}

\restylefloat{figure}

\begin{document}

\title[ ]{\emph{RXTE} Timing Analysis of the AXP 1E 2259+586}
\author[B. \.{I}\c{c}dem et al.]
{B.~\.{I}\c{c}dem,$^1$ A.~Baykal,$^1$ and S. \c{C}.~\.{I}nam,$^2$\\
$^1$ Physics Department, Middle East Technical University, Ankara, Turkey\\
$^2$ Department of Electrical and Electronics Engineering, Ba\c{s}kent University, Ankara, Turkey}

\date{Received \line(1,0){60}/ Accepted \line(1,0){60}}

\maketitle

\begin{abstract}

We present pulse timing analysis of \emph{Rossi X-ray Timing Explorer}
(\emph{RXTE}) observations of the anomalous X-ray pulsar 1E 2259+586 
from its 2002 outburst to October 2010.
Our objectives are to extend
the work on the recovery stage after the 2002 glitch, investigate the
variations caused by the second glitch which occurred in 2007 and look for other
unusual events, if any, that arise in the regular spin-down trend of the
source. We find that the fractional change in the spin frequency derivative 
after the 2002 glitch
is not stable since it decreased an order of magnitude,
 from $-2.2\times10^{-2}$ to $-1.278(3)\times10^{-3}$,
in about 2.5 years. From pulse timing analysis, we discover
 two small frequency shifts with fractional changes
 $\Delta\nu/\nu=3.08(32)\times10^{-8}$ and
$\Delta\nu/\nu=-1.39(11)\times10^{-8}$. While the first one of these shifts 
is not found to have a fractional frequency derivative change while
second one has $\Delta\dot\nu/\dot\nu=-2.9(2)\times10^{-2}$.
We interpret these frequency changes as positive and negative
microglitches similar to those seen in radio pulsars.

\end{abstract}


\section{Introduction}

The X-ray pulsar (AXP) 1E 2259+586 was discovered by \citet{fah81} 
within the supernova remnant G109.1-1.0 with $\sim7$ s pulsations.
Further X-ray observations of 1E 2259+586 with Tenma, EXOSAT and Ginga 
\citep{koy87,koy89,mor88,han88,iwa92} 
have indicated that the AXP is steadily spinning down with a rate of $\dot\nu\sim-1.0\times10^{-14}$ Hz.s$^{-1}$,
which is too slow to power the observed X-ray luminosity.
Using \emph{ROSAT} observations,
 \citet{bay96} found a marginal spin-up episode between 1991 and 1992.
\citet*{mer98} using the RXTE observations 
gave an upper limit on the semi-axis to be 0.03 lt.sec suggesting a white dwarf companion smaller than 0.8 M$_\odot$ (see also \citet{bay98}).
Using RXTE observations, 
\citet*{kas99} obtained, for the first time, a phase coherent timing solution for 1E 2259+586 , indicating 
stability over a 2.6 yr time period. This stability strongly 
suggested the
 magnetar interpretation rather than the binary accretion hypothesis.
 Using ASCA and BBXRT observations, 
\citet{cor95} showed for the first time that the X-ray spectra can be explained by
a blackbody plus power law model. INTEGRAL and RXTE observations have revealed
the evidence of hard spectral tails above 10 keV \citep{kui06} 
(see also \citet{kas10}).

In June 2002, 1E 2259+586 showed an outburst consisting of over 80 short SGR-like  bursts 
\citep*{gav04}. During this outburst, flux level of the source increased by a factor of 
more than 20 accompanied with a glitch of fractional frequency change 
$\frac{\Delta\nu}{\nu}\sim 4.24\times10^{-6}$ 
\citep{kas03,woo04}. In 2007, a second glitch of $\frac{\Delta\nu}{\nu}\sim 8.5\times10^{-7}$
was observed when the source was quiescent \citep*{dib08}.

\citet{woo04} modelled  \emph{XMM} and \emph{RXTE} spectra of 1E 2259+586
before and after the 2002 outburst with the blackbody 
plus power law model modified by interstellar absorption. 
The blackbody temperature increased rapidly after the outburst in contrast 
to the photon index which decayed about half of its pre-outburst value. Both parameters 
quickly recovered to within $25\%$ of their their pre-burst levels
within the first $\sim$1--3 days \citep{woo04}. 
In the first few hours of the outburst, \emph{XMM} and \emph{RXTE} observations revealed that
pulsed fraction decreased to $\sim15\%$ then, like the spectral parameters, 
it quickly recovered to its pre-outburst state within 6 days. 
There were also some variations in the phase dependence of the
energy spectrum: While the photon index showed significant variability one
week before the outburst, this variation vanished three days after the outburst.
XMM-Newton observations of the source showed that 
 X-ray flux decayed with a power law index $\sim -0.69 $ after 2002 outburst
until 2005. A strong correlation between X-ray flux and hardness ratio is 
observed by \citet{zhu08}.
Pulse profile of the source changed in such a way that much of the power moved to the
fundamental harmonic and the second harmonic regained its power within 1 week
after which a very slow recovery  within several weeks was observed \citep{woo04}. 
Using the cumulative properties of the outburst, \citet{woo04} concluded that the pulsar experienced
something which was not sudden at all and including two components,
one of them was on the surface such as a series of fractures and the other distributed over a wider
region like smoother plastic change. 
Hence, both the superfluid interior and the magnetosphere were affected.

The \emph{Spitzer Space Telescope} observations of \citet{kap09} at near- and mid-infrared
regions of the electromagnetic spectrum was interpreted as a possible indication of the presence of a passive
X-ray-heated dust disc. \citet{kap09} also showed that the IR data of this source can also be modelled
 by a power law spectrum which could be explained by
magnetospheric emission. \citet*{tia10} estimated the distance
to the AXP (and also to the SNR G109.1-1.0) as $4.0\pm0.8$ kpc using the 21-cm H{\small \ I}-line
and CO-line spectra of the SNR G109.1-1.0, H{\small \ II} region Sh 152, and the adjacent
molecular cloud complex.

In this paper, we construct phase coherent pulse arrival times between 2002 and 2010.
Using these arrival times we refine the timing solution after the 2002 glitch 
and look for frequency and derivative changes in these time intervals.

\section{Data \& Analysis}
\subsection{Data}

The dataset used in this work consists of \emph{Rossi X-Ray Timing Explorer} 
(\emph{RXTE}) proportional counter array (PCA) observations of the AXP 1E 2259+586 
covering the time between March 2000 (MJD 51613) and October 2010 (MJD 55483). 
The PCA operates in the energy range 2--60 keV using an array of five collimated
xenon/methane multi-anode proportional counter units (PCUs). The instrument 
has a total effective area of $\sim6500$ cm$^2$ and a field of view of $\sim1^o$
FWHM \citep{jah96}. 

A total of 479 observations were used for the analysis presented in this paper.
The durations of observations vary from 10.6 ks to 0.2 ks, most of them being
greater than 1 ks.
 During the analysed \emph{RXTE}-PCA observations, the number of active PCUs varied
between 1 and 4. Due to timing concerns, we used all the available layers of all PCUs 
in our analysis. 

\subsection{Pulse-Timing Analysis}

The GoodXenonWithPropane or GoodXenon mode 
and event mode data were energy selected (2--10 keV) for all xenon layers, and
were binned with 125 ms time resolution. 
The time values in the light curves were then corrected to the solar system barycentre.
We used the standard analysis tools for \emph{RXTE}-PCA data included in FTOOLS package
to obtain the light curves and to merge them into a single light curve covering all the interval
mentioned above. In the extraction of lightcurves we removed all outburst events in order to avoid 
sudden pulse profile changes. The light curves were folded at frequencies given by \citet{woo04}.
The time series 
was split into intervals of approximately same duration,
 and each segment was folded with a quadratic ephemeris 
with the same frequency and frequency derivative,
so that we obtained a pulse profile for each time interval, which
is made up of 20 phase bins. Then, we switched to the harmonic representation of pulse profiles 
as introduced by \citet{boy85}

\begin{eqnarray}
 f(\phi) &=& F_0+\sum_{k=1}^{10}F_k\cos k(\phi-\phi_k), 
\end{eqnarray}
The template pulse is extracted from a longer time interval at which glitch recovery 
has already taken place

\begin{eqnarray}
  g(\phi) &=& \sum_{k=1}^{10}G_k\cos k(\phi-\phi_k) . \\
\end{eqnarray}
Then, 
by cross-correlating the pulse profiles with the template pulse, 
 we obtained the pulse arrival times, $\Delta\phi$. 
 In order to see the effect of pulse profile changes in pulse timing, we also 
 performed pulse 
timing using first and second harmonics independently. Our pulse arrival times 
were consistent within 1$\sigma $ level with 10 harmonic expansion of pulse profiles
for 20 phase bins. This suggests that even if the pulse profile changes, pulse 
phases are not effected. As a final check, we gave weight to all pulse harmonics 
using the variances of harmonics from all pulses. We applied the cross correlation 
for the weighted pulses (or filtered pulses), and again we obtained the pulse arrival times as 
those of unweighted pulses in 1$\sigma$ level confidence. We concluded that pulse shape changes do not 
have any effect on pulse phases and pulse timing analysis. 

\begin{figure*}[htb]
\begin{center}
\subfloat{
\includegraphics[trim=0cm 0cm 1.5cm 0cm, clip=true, angle=0, width=0.8\textwidth,totalheight=0.75\textheight]{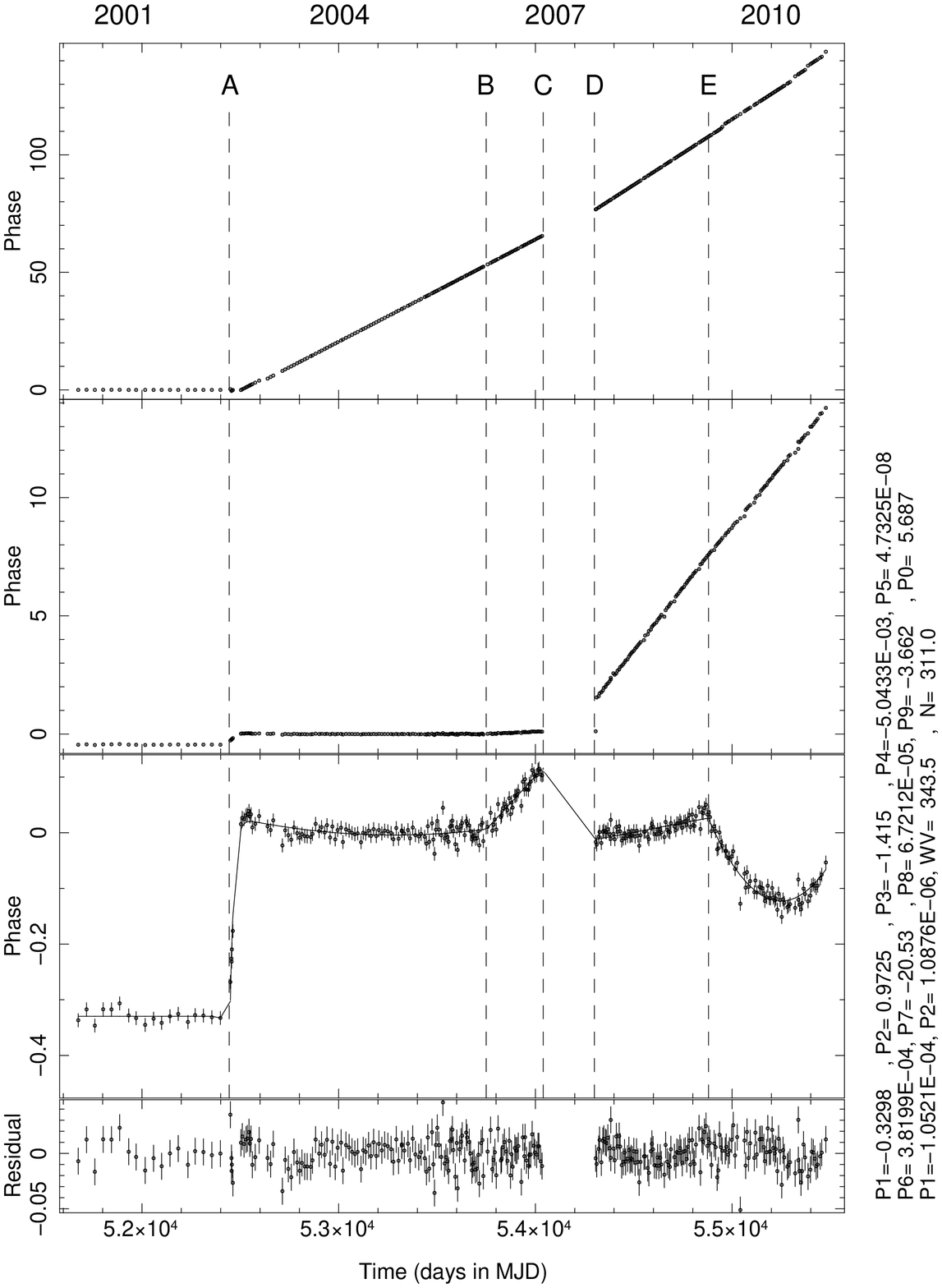}} 
\end{center}
\caption{Phase offset series for AXP 1E 2259+586. Panels are described from top
to bottom. \emph{Top panel}:
 Phase offsets extracted using the spin-down model of the period before 2002-glitch.
\emph{Second panel}:
 Phase offsets using the correction of the pulse frequency after MJD 53443.13.
\emph{Third panel}:
 Arrival times obtained by an additional correction after MJD 54300. The solid curve 
is the phase offset evolution of the models presented in Tables \ref{tab:analysis_glitch1} and \ref{tab:analysis_glitch3}.
\emph{Bottom panel}:
 Residuals, after subtracting the best-fitting models given in Tables 
\ref{tab:analysis_glitch1} and \ref{tab:analysis_glitch3}.
\emph{All panels}: Glitch epochs are indicated with dashed vertical lines. A: 2002 glitch
corresponds to MJD 52443.13, B: small microglitch at MJD $\sim$53750, C-D: no phase-connection, 
D: second glitch at MJD  $\sim$54300,
E: small microglitch at MJD $\sim$54880.}
\label{fig:glitch}
\end{figure*}

In top panel of Fig. \ref{fig:glitch}, it is seen that 
the slope of the pulse arrival times changes after the first glitch.
In the estimation of pulse arrival times, the spin down rate 
of the source was kept fixed at the value before the glitch, 
which is $\dot\nu = -9.92\times 10^{-15}$ Hz.s$^{-1}$.
By modelling the pulse arrival times lying after the first glitch  
by a first order polynomial, 
with $\Delta\phi=\phi_0+\delta\nu t$,
the new pulse arrival times were obtained for the corrected pulse frequency, 
$\nu_0+\delta\nu$. In the second panel of Fig. \ref{fig:glitch}, it 
is clearly seen that the slope of the pulse 
arrival times is shifted again after MJD 54300.
The shift in the pulse frequency is found again using the method explained above
and new pulse arrival times are obtained after MJD 54300 by correcting the pulse 
frequency again.
The third panel of Fig. \ref{fig:glitch}, where
 we present pulse arrival times for the corrected frequencies 
after the first and second glitches, shows
 that there are additional breaks in pulse arrival time 
series around MJD 53750 and 54880 (see vertical dash lines quoted B and E)

The first glitch is at MJD 52443.13 and was reported
by \citet{kas03} to be accompanied by an outburst of many individual bursts.
About 1360 days after the first glitch,  
 at around MJD 53750, we resolved 
a small fractional frequency shift of $\frac{\Delta\nu}{\nu}= 3.08(32)\times10^{-8}$.
 Phase connected pulse arrival times after MJD 54300  indicates 
another glitch with a fractional
 frequency change of $\frac{\Delta\nu}{\nu}\sim8.20(2)\times10^{-7}$.
This second glitch is consistent 
with 1 $\sigma $ level given by \citet{dib08} in their review paper of AXP glitches.
Moreover, around MJD 54880, we have observed another 
fractional frequency shift of $\frac{\Delta\nu}{\nu}= -1.39(11)\times10^{-8}$ with 
$\frac{\Delta\dot\nu}{\dot\nu}=-0.029(2)$.
All the glitch episodes in this work
 are presented in Fig. \ref{fig:glitch} and timing solution are given in 
Tables \ref{tab:analysis_glitch1} and \ref{tab:analysis_glitch3}.

For the first glitch,
as described in \citet{woo04}, the standard post-glitch relaxation models, 
which consist of single or multiple exponential relaxation terms (Alpar et al. 1984b),
did not provide reasonable fits.
An alternative model is the irregular model including an exponential rise
term, which had been developed for two glitches of Crab pulsar \citep*{lyn93, won01}.
 In this model, the increase 
in the slowdown rate is interpreted as the cumulative of successive glitches;
part of the increase is provided by a step and part is in the form of 
an exponential rise 
\\
\begin{eqnarray}
\nu=\nu_0+\dot{\nu}_0(t-t_0)+\Delta\nu+\Delta\nu_g(1-e^{-(t-t_g)/\tau_g}) \nonumber \\
-\Delta\nu_d(1-e^{-(t-t_g)/\tau_d})+\Delta\dot{\nu}t,
\end{eqnarray}
where the first two terms represent the frequency evolution before the glitch,
$t_g$ is the glitch epoch, 
$\Delta\nu$ is the frequency jump with the glitch, $\Delta\nu_g$ and $\Delta\nu_d$ are
the growth and decay amplitudes,
 respectively, $\tau_g$ and $\tau_d$ are the growth and decay time-
scales respectively, and $\Delta\dot\nu$ is the jump in the spin down rate.
We used the corresponding phase evolution
 equation to fit our pulse arrival times in the period MJD 52390--53750
(The correction to the pulse frequency after the first glitch mentioned above is 
$\delta\nu=\Delta\nu+\Delta\nu_g-\Delta\nu_d$).
In the fitting process, we used the $\chi^{2}$ statistic and leave 
all parameters as free parameter. For normalizations of exponentials 
we used the initial ranges as given by \citet{woo04}.

  In Table \ref{tab:analysis_glitch1}, we present timing solution
of the first glitch for the extended time coverage 
from MJD 51613 to MJD 53750. 
Our analysis indicates that $\frac{\Delta\dot\nu }{\dot\nu}$
 has shown a variation from $-2.2\times10^{-2}$ to $-1.278(3)\times10^{-3}$,
i.e. there is an order of magnitude reduction in the fraction in a $\sim2.5$-year time.
We also fitted the pulse arrival times for a time interval given by \citet{woo04} and found that 
the timing solution is consistent with Woods et al. (2004) in 1 $\sigma$ level.

The small frequency shifts of $\frac{\Delta \nu}{\nu} \sim 10^{-8}$ with both signs 
at MJD 53750 and 54880 are at the order of microglitches seen 
in radio pulsars. In the first frequency shift, $\frac{\Delta \dot \nu }{\dot \nu} $ is not significant, 
however in the second one this value is found to be $-0.029(2)$, i.e.  
the fractional change in the pulse frequency derivative is negative.
This is not very unusual when compared to the microglitches in radio pulsars,
since there is no preferred sign for both of the fractional jumps 
$\Delta\nu/\nu$ and $\Delta\dot\nu/\dot\nu$ \citep{chu10}. 
 The timing parameters of these small
 frequency shifts are given in Table \ref{tab:analysis_glitch3}.

\begin{table*}
\caption{Timing Solution for 1E 2259+586 Before MJD 53750}
\begin{tabular}[htb]{l c c}
\hline 
\hline
\multirow{2}{*}{Parameter} &
\multirow{2}{*}{Value $^1$} &
\multirow{2}{*}{Value $^2$} 
\\ 
& &
\\
\hline
\\
{Spin frequency,$\nu$ (Hz)} & {$0.14328703257(21)$} 
& {$0.14328703257(21)$} 
\\
{Spin frequency derivative, $\dot{\nu}$ (Hz.s$^{-1}$)} 
& {$-9.920(6)\times10^{-15}$} & {$-9.920(6)\times10^{-15}$}  
\\
{Epoch (MJD)} & {$52400$} 
& {$52400$} 
\\
{$\Delta\nu$ (Hz)} & {$5.25(12)\times10^{-7}$} 
& {$6.70(1.14)\times10^{-7}$}  
\\
{$\Delta\nu_g$ (Hz)} & {$>8.7\times10^{-7}$} 
& {$8.29(78)\times10^{-7}$} 
\\
{$\tau_g$ (days)} & {$14.1(7)$} 
& {$14.1(1.2)$} 
\\ 
{$\Delta\nu_d$ (Hz)} & {$\Delta\nu_g+(\sim5\times10^{-9})$} 
& {$1.06(8)\times10^{-6}$} 
\\ 
{$\tau_d$ (days)} & {$15.9(6)$} 
& {$15.9(1.1)$} 
\\ 
{$\Delta\dot{\nu}$ (Hz.s$^{-1}$)} & {$2.18(25)\times10^{-16}$} 
& {$1.268(3)\times10^{-17}$} 
\\ 
{$t_g$ (MJD)} & {$52443.13(9)$} 
& {$52443.13(10)$} 
\\ 
{rms timing residual (ms)} & {$44.9$} 
& {85.1} 
\\ 
{Start observing epoch (MJD)} & {$51613$} 
& {$51613$} 
\\ 
{End observing epoch (MJD)} & {$52900$} 
& {$53750$} 
\\ 
\hline
\end{tabular} \\
Numbers in parentheses indicate 1$\sigma$ uncertainties in the least significant digits quoted.
$^1$ Taken from table 4 of \citet{woo04}. $^2$ From our analysis.
\label{tab:analysis_glitch1}
\end{table*}

\begin{table*}
\caption{Timing Solution for 1E 2259+586 After MJD 53750}
\begin{tabular}[htb]{l c c c}
\hline
\hline
\multirow{2}{*}{Parameter} &
\multirow{2}{*}{Microglitch 1} &
\multirow{2}{*}{Glitch 2} &
\multirow{2}{*}{Microglitch 2}  
\\ 
& 
\\
\hline
\\
{Spin frequency,$\nu$, (Hz)} & {$0.143286381(13)$} & {$0.143286138(14)$} & {$0.143285760(14)$} 
\\
{Spin frequency derivative, $\dot{\nu}$ (Hz.s$^{-1}$)} & {$-9.920(6)\times10^{-15}$} & 
{$-9.920(6)\times10^{-15}$} & {$-9.920(6)\times10^{-15}$}
\\
{Epoch (MJD)} & {53750} & {54040} & {54880}
\\
{$\Delta\nu$ (Hz)} & {$4.42(46)\times10^{-9}$} & {$>1.6\times10^{-6} $} & {$-2.00(15)\times10^{-9}$}
\\ 
{$\Delta\dot{\nu}$ (Hz.s$^{-1}$)} & {--} & {--} & {$2.91(22)\times10^{-16}$}
\\ 
{$t_g$ (MJD)} & {$\sim53750$} & {$\sim54040$} & {$\sim54880$}  
\\ 
{rms timing residual (ms)} & {76.8} & {66.8} & {106.7}  
\\ 
{Start observing epoch (MJD)} & {53700} & {53900} & {54800}  
\\ 
{End observing epoch (MJD)} & {54040} & {54900} & {55570} 
\\ 
\hline
\end{tabular}\\
Numbers in parentheses give 1$\sigma$ uncertainties in the least significant digits quoted.
\label{tab:analysis_glitch3}
\end{table*}

\section{Discussion}

In this work we presented  extended pulse timing analysis of 1E 2259+586 from 2002 to 2010.
In phase coherent pulse timing analysis, we found that after 2002 glitch 
$\frac{\Delta\dot\nu }{\dot\nu}$ reduced an order of magnitude  
in a $\sim2.5$-year time.
We also discover two micro glitches  
$\Delta\nu/\nu=3.08(32)\times10^{-8}$ and $\Delta\nu/\nu=-1.39(11)\times10^{-8}$.
In the first frequency shift, we do not resolve
frequency derivative.
 In the latter one, fractional change in the spin down rate is found to be
 $\Delta\dot\nu/\dot\nu=-2.9(2)\times10^{-2}$. 

Microglitches make up a class of small amplitude resolvable jumps that can be observed in both, or
either of pulsar rotation frequency or its first time derivative \citep{cor85,chu02,cor88,ale95}.
The orders of magnitude for $\nu $ and $\dot \nu $ are in the range
$10^{-11} < |\Delta \nu / \nu| <  10^{-8}$ and
$10^{-5} < |\Delta \dot \nu / \dot \nu | < 10^{-2} $, respectively.
So far 319 micro glitches are observed from 46 radio pulsars
\citep{cor85,ale95,chu10}.
Microglitches observed in 1E 2259+586 are of the order of 
the upper limits of radio pulsar microglitches in magnitude . 
Unlike the large glitches, microglitches can have both negative and positive signs in 
their fractional pulse frequencies and first time derivatives. 
Hence,it is difficult to understand microglitches in terms of theories developed for 
large glitches \citep{cor85,cor88}. 
 The cause of micro glitches is unclear but can be explained as
involving neutron star interior and/or magnetospheric torque fluctuations
\citep{rud98}. The magnetar crust can be cracked
by the variations of the magnetic field
configuration \citep{rud98}. In this case, both signs of microglitches may be observed. 

Anti-glitch (negative glitch) behaviour  
have been seen in two other magnetars \citep{tho00,gav11}. Sudden spin down trend of  
SGR 1900+14 is interpreted by \citet{tho00} as an anti-glitch with a fractional pulse frequency change
of $\Delta \nu / \nu\sim 10^{-4}$. However, the scale of this event is 
four orders of magnitude larger than that of the negative microglitch of 1E 2259+586. 
For AXP 4U 0142+61, 
\citet*{gav11} showed that
the total fractional frequency jump at the time of 
glitch is $\Delta \nu / \nu\sim 1.7 \times 10^{-6}$; however, the decay of large 
glitch event left a net negative 
slow-down portion of $\Delta \nu / \nu\sim -8  \times 10^{-8}$ after 17 days of recovery.  
 In addition, for rotation powered pulsar PSR J1846-0258, \citet*{liv10} 
 reported a large spin up glitch which experienced a large recovery and ended up with a net
 decrease of pulse frequency similar to the recovery of AXP 4U 0142+61 and event 
occurred in SGR 1900+14. 
So far 1E 2259+586 is the unique source showing microglitch/anti-glitch events among the AXPs and SGRs
\citep{dib08}. 

In magnetars, glitches may be triggered by strong internal magnetic
 fields as the crust is deformed, either plastically or cracked violently \citep{tho96}.
 There may be some parts of superfluid rotating more slowly
than the crust. Then, by a glitch the angular momentum is transferred from the crust 
to the superfluid, opposing the case in radio pulsar glitches,
and the crust spins down after the glitch.
This explanation is proposed by \citet{tho00} to explain the net
 spin down rate event in SGR 1900+14. 
Hence, small scale magnetar glitches can explain the glitch/anti-glitch trend of micro glitches  
like those seen in 1E 2259+586.
 
The 2002 glitch is well-studied by \citet{woo04} until the post glitch recovery 
mid of 2003.  We extended post glitch recovery until the first micro glitch event seen in 2006.  
In the pulse timing
analysis, we confirmed that the fractional frequency change of the glitch has an amplitude of
$\Delta\nu/\nu=4.68(80)\times10^{-6}$, and other post glitch parameters (see Table 1)
  determined by \citet{woo04} from their
500-day time span of observations. The fractional change in the pulse frequency derivative
$\Delta\dot\nu/\dot\nu=-1.278(3)\times10^{-3}$, from our 1360-day pulse arrival times
is found to be 17 times smaller than that of \citet{woo04}, which is deduced from
a shorter time interval.  
 We should note that our analysis in their timing interval yielded a similar value
in $1-\sigma$ limit. 
(This $1-\sigma$ consistency in $\Delta\dot\nu/\dot\nu $
 does not disrupt even if we freeze the normalizations of the exponential components 
to the values given by \citet{woo04}).  
 This indicates that fractional change of pulse frequency derivative is variable.
As discussed in  \citet{woo04} the negative sign of the fractional change of derivative of
pulse frequencies is quite unusual for large radio pulsar glitches. 
For the large glitches fractional frequency change is $\Delta\nu/\nu>10^{-7}$ and 
fractional pulse frequency derivative
has the range $\Delta\dot\nu/\dot\nu\sim 10^{-2}-10^{-3}$ (see also \citet{alp94,alp06}). 

In the standard 
large glitches of radio pulsars, 
the sudden shifts in the pulse frequency and the spin down rate of a radio pulsar have been considered 
to be due to the coupling mechanism between
 the crust and the crust superfluid (Alpar et al. 1984a,b); \citep*{alp89}. 
The angular momentum transfer
 between the crust superfluid and the crust is provided by the quantized vortices. 
There are two types of vortex lines separated according to their dynamical behaviour in the
superfluid. One type is responsible for the continuous vortex flow since they are
not pinned to the inner crust of the neutron star but they unpin and repin at some rate.
The other type of vortices are pinned to the crustal nuclei and are not allowed to move radially
outwards up to the maximum tolerance of the pinning force.
 Hence, a lag appears between the spin-down rates 
of the crust and the superfluid, $\nu=\nu_s-\nu_c > 0$.
 When the vortices cannot stand this lag any more, 
they suddenly unpin from the crustal nuclei and some angular momentum is transferred to
the crust. This sudden unpinning of the superfluid neutron vortices from the crustal nuclei can result in
glitches of fractional pulse frequency changes of $\Delta\nu/\nu\sim 10^{-7}$ to $10^{-6}$ 
together with increases in the magnitudes of spin down rate. The latter type of vortices
do not have any contribution to the usual spin down trend of the pulsar other than the 
glitches. Hence, the relaxation of the spin down rate after the glitch is accomplished
by the continuous vortex flow, and the persistent change in the spin down 
rate, which is observed sometimes, is interpreted as the increase in the amount of 
the pinned vortices that cannot move towards the outer crust after the glitch.
If the superfluid part supplying the vortex flow has moment of inertia $I_A$ and 
the trapped part has the moment of inertia $I_B$, then the equations governing the 
dynamics of the glitches are determined as follows \citep*{alp95}
\begin{eqnarray}
 \frac{\Delta\nu}{\nu} &=& \left(\frac{I_A}{2I}+\frac{I_B}{I}\right)\frac{\delta\nu}{\nu}, \\
\frac{\Delta\dot\nu}{\dot\nu} &=& \frac{I_A}{I},
\end{eqnarray}
where $I$ is the total moment of inertia of the pulsar and $\delta\nu$ is the
change of the rotation frequency of the superfluid interior. Thus, in the vortex creep theory
the fractional change in the
spin down rate is determined by the fraction $I_A/I$ which should have a positive value
in the presence of a constant external torque applied on the pulsar.
Unless the vortices propagate inwards to the rotation axis of the star, 
the negative value of $\frac{\Delta\dot\nu}{\dot\nu}$ can be explained by external torques.
This could be due to the secular change of magnetic
 moment \citep{rud91} or change in the electromagnetic torque by a reorganization of magnetic moment 
and angular momentum axis of the star \citep{link97}. However, 
fractional change of the
pulse frequency derivative by a factor of 17 in a time scale of only 860 days due to the 
magnetic moment change is quite fast. 
Another  explanation could be relaxation of twisted magnetic fields which creates stresses on the 
crust \citep{tho02}. Persistent 
seismic activity may modulate the spin down torque and crustal shear waves drives particles wind and 
causes angular momentum changes \citep{tho98,har99,tho00}. 
It is interesting to note that all glitches of 1E 2259+586 have
occurred almost at a constant spin down rate which is $\dot \nu = -9.920 \times 10^{-15}$ Hz $^{-1}$. 
This suggests that glitches happen to be when post glitch frequency derivatives are completely relaxed. 
Therefore, the excitation of precession during the glitch, due to a stable reorientation 
of magnetic axis relative the rotation axis of the star is considered unlikely since 
long term torque on the star is not stable \citep{link97}. 

As a final remark, after 2002 glitch,
 \citet{kas03} report $\Delta \dot \nu / \dot \nu \sim 1.1 $ for a time span
of a couple of weeks. They suggest that the increase of spin down rate is due to 
the decoupling of crust and core superfluid
where they recouple in 2 weeks time-scale which is extremely long compared to the
theoretical estimates \citep{alp88,sid09}. This is also clearly
 resolved in our pulse timing analysis for a particular short time span; 
however, as we are interested in new events in the
 long term recovery and unusual events in pulse timing, this aspect
of 2002 glitch is kept out of scope of this work.

\section*{acknowledgments}
We acknowledge research project TBAG 109T748 of the Scientifi�c and Technological
Research Council of Turkey (T\"{U}B\.{I}TAK). A.B. also appreciate the useful discussions
with M. Ali Alpar and  
K.S.Cheng for the comment on the paper. We thank anonymous referee for the useful suggestions to
improve the paper.

\bibliographystyle{mn2e}
\bibliography{2259_ref_mnras11}

\end{document}